%% file: ad_insertion_icmi.tex
\documentclass[sigconf,screen]{acmart}

\usepackage{booktabs} 
\usepackage{amsfonts}
\usepackage{amssymb}
\usepackage{xspace}
\usepackage{epstopdf}
\usepackage{enumitem}
\usepackage{algorithm}
\usepackage{algorithmic}
\usepackage{multirow}
\usepackage{balance}
\usepackage{booktabs}
\usepackage{array}
\usepackage{wrapfig}
\usepackage{pifont}
\usepackage{pbox}
\usepackage{geometry}
\usepackage{color}

\setcopyright{acmlicensed}
\acmPrice{15.00}
\acmDOI{10.1145/3136755.3136796}
\acmYear{2017}
\copyrightyear{2017}
\acmISBN{978-1-4503-5543-8/17/11}
\acmConference[ICMI'17]{19th ACM International Conference on Multimodal Interaction}{November 13--17, 2017}{Glasgow, UK}

\renewcommand\footnotetextcopyrightpermission[1]{} 

\begin{document}

\def\x{{\mathbf x}}
\def\L{{\cal L}}
\def\x{{\mathbf x}}
\def\L{{\cal L}}
\def\eg{\textit{e.g.}}
\def\ie{\textit{i.e.}}
\def\Eg{\textit{E.g.}}
\def\etal{\textit{et al.}}
\def\etc{\textit{etc.}}
\def\nback{\textit{n}-back }
\setlength{\tabcolsep}{2pt}

\title{Evaluating Content-centric vs User-centric Ad Affect Recognition}

\author{Abhinav Shukla}
\affiliation{%
  \institution{International Institute of Information Technology}
  \city{Hyderabad} 
  \country{India} 
}
\email{abhinav.shukla@research.iiit.ac.in}

\author{Shruti Shriya Gullapuram}
\affiliation{%
 \institution{International Institute of Information Technology}
  \city{Hyderabad} 
  \country{India} 
}
\email{shruti.gullapuram@students.iiit.ac.in}

\author{Harish Katti}
\affiliation{
  \institution{Centre for Neuroscience, Indian Institute of Science}
  \city{Bangalore} 
  \country{India}
}
\email{harish2006@gmail.com}

\author{Karthik Yadati}
\affiliation{%
  \institution{Delft University of Technology}
  \city{Delft}
  \country{Netherlands}
}
\email{n.k.yadati@tudelft.nl}

\author{Mohan Kankanhalli} 
\affiliation{%
 \institution{School of Computing, National University of Singapore}
 \city{Singapore} 
 \country{Singapore}}
\email{mohan@comp.nus.edu.sg}

\author{Ramanathan Subramanian}
\affiliation{%
  \institution{School of Computing Science, University of Glasgow}
  \city{Singapore} 
  \country{Singapore}
}
\email{ramanathan.subramanian@glasgow.ac.uk}

\renewcommand{\shortauthors}{A. Shukla et al.}

\begin{abstract}
Despite the fact that \textbf{advertisements} (ads) often include strongly emotional content, very little work has been devoted to affect recognition (AR) from ads. This work explicitly compares \textbf{content-centric} and \textbf{user-centric} ad AR methodologies, and evaluates the impact of enhanced AR on computational advertising via a user study. Specifically, we (1) compile an affective ad dataset capable of evoking coherent emotions across users; (2) explore the efficacy of \textit{content-centric} convolutional neural network (CNN) features for encoding emotions, and show that CNN features outperform low-level emotion descriptors; (3) examine \textit{user-centered} ad AR by analyzing Electroencephalogram (EEG) responses acquired from eleven viewers, and find that EEG signals encode emotional information better than content descriptors; (4) investigate the relationship between objective AR and subjective viewer experience while watching an ad-embedded online video stream based on a study involving 12 users. To our knowledge, this is the first work to (a) expressly compare user vs content-centered AR for ads, and (b) study the relationship between modeling of ad emotions and its impact on a real-life advertising application.  
\end{abstract}

\ccsdesc[500]{Human-centered computing~HCI theory, concepts and models}
\ccsdesc[300]{Human-centered computing~User centered design}


\keywords{Affect recognition, Ads, Content-centric vs User-centric, CNNs, EEG, Multimodal analytics, Computational Advertising}

\maketitle

\section{Introduction}~\label{sec:intro}
Advertising is a rapidly evolving global industry that aims to induce consumers into preferentially buying specific products or services. In this digital age, audio-visual content is increasingly becoming the preferred means of delivering advertising campaigns. The global advertising industry is estimated to be worth over US \$500  billion\footnote{http://www.cnbc.com/2016/12/05/global-ad-spend-to-slow-in-2017-while-2016-sales-were-nearly-500bn.html}, and web advertising is expected to be a key profit-making sector with video advertising playing a significant role\footnote{http://www.pwc.com/gx/en/industries/entertainment-media/outlook/segment-insights/internet-advertising.html}. Advertisements (ads) often contain strongly emotional content to convey an effective message to viewers. Ad \textit{valence} (pleasantness) and \textit{arousal} (emotional intensity) are key properties that modulate emotional values and consumer attitudes associated with the advertised product~\cite{Holbrook1984,Holbrook1987,Pham2013}. In the context of Internet video advertising (as with \textit{YouTube}), modeling the emotional relevance between ad and program content can improve program comprehension and advertisement brand recall, as well as optimize user experience~\cite{cavva}. 

Even though automated mining of ad emotions is beneficial, surprisingly very few works have attempted to computationally recognize ad emotions. This is despite the field of \textit{\textbf{affective computing}} receiving considerable interest in the recent past, and a multitude of works modeling emotions elicited by image~\cite{katti2010making,maneesh2017acii}, speech~\cite{lee2005toward}, audio~\cite{AAAI17}, music~\cite{Koelstra} and movie~\cite{decaf,subramanian2016ascertain} content. Overall, affect recognition (AR) methods can be broadly classified as \textit{content-centric} or \textit{user-centric}. \textit{Content-centric} AR approaches characterize emotions elicited by multimedia content via textual, audio and visual cues~\cite{Hanjalic2005,wang2006affective}. In contrast, \textit{user-centric} AR methods aim to recognize the elicited emotions by monitoring the user or multimedia consumer via facial~\cite{joho2011looking} or physiological~\cite{Koelstra,decaf,subramanian2016ascertain,Subramanian2014} measurements.  

This paper expressly examines and compares the utility of \textit{content-centric} and \textit{user-centric} approaches for ad AR. As emotion is a subjective human feeling, most recent AR methods have focused on a variety of human behavioral cues. Nevertheless, ads are different from conventional media such as movies, and are compact representations of themes and concepts which aim to impact the viewer within a short span of time. Thus, it would be reasonable to expect that ads contain powerful audio-visual content to convey the intended emotional message. While some works have compared content and user-centric features for AR, an explicit comparison has not been performed for ads to our knowledge. Another question that we try to answer in this work, perhaps for the first time in affective computing, is \textit{whether improved AR as given by objective measures, directly impacts subjective human experience} while using a multimedia application.

We first present a carefully curated affective ad dataset, capable of evoking coherent emotions across viewers as seen from emotional impressions reported by \textit{experts} and \textit{novice annotators}. On ensuring that the ads are able to reliably evoke target emotions (in terms of arousal and valence levels), we examine the efficacy of content and user-based methods for modeling ad emotions-- specifically, high-level convolutional neural network (CNN) features and low-level audio visual descriptors~\cite{Hanjalic2005} are explored for content-centered analysis, while EEG measurements are employed for user-centered AR. CNN features outperform low-level audio-visual descriptors, but are inferior to EEG signals implying that user-centric cues enable superior ad AR. We then show how improved AR achieved by the CNN and EEG features reflects in terms of better ad memorability and user experience for a computational advertising application~\cite{cavva}. 

To summarize, this work makes the following contributions: (1) To our knowledge, this is the first work to explicitly compare and contrast content-centered and user-centered ad AR; (2) This is also the first work to demonstrate how an improvement in objective AR performance improves subjective ad memorability and user experience while watching an ad-embedded online video stream. Our findings show that enhanced AR can facilitate better ad insertion onto broadcast multimedia content; (3) The compiled dataset of 100 affective ads along with accompanying subjective ratings and EEG responses is unique for ad-based AR. 

The paper is organized as follows. Section~\ref{RW} reviews related literature, while Section~\ref{ad_set} overviews the compiled ad dataset and the EEG acquisition protocol. Section~\ref{Data_anal} presents the techniques adopted for content and user-centered ad AR, while Section~\ref{ER} discusses AR results. Section~\ref{US} describes a user study to establish how improved AR facilitates computational advertising. Section~\ref{CFW} summarizes the main findings and concludes the paper.

\section{Related Work}\label{RW}
To position our work with respect to the literature and highlight its novelty, we review the related work examining (a) affect recognition (b) the impact of affective ads on consumer behavior (c) computational advertising.

\subsection{Affect recognition}
Building on the circumplex emotion model that represents emotions in terms of valence and arousal~\cite{Russell1980}, many computational methods have been designed for affect recognition. Typically, such approaches are either \textit{content-centric} which employ image, audio and video-based emotion correlates~\cite{Hanjalic2005,vonikakis2017probabilistic,Shukla2017acm} to recognize affect in a supervised manner; or \textit{user-centric}, which measure stimulus-driven variations in specific physiological signals such as pupillary dilation~\cite{YadatiMMM2013}, gazing patterns~\cite{Subramanian2014,Tavakoli15} and neural activity~\cite{Koelstra,decaf,Zheng2014}. Performance of these models is typically subject to the variability in subjective, human-annotated labels, and careful affective labeling is crucial for successful AR. We carefully curate a set of 100 ads such that they are assigned very similar emotional labels by two independent groups comprising experts and novice annotators. These ads are then mined for emotional content via content and user-based methods. User-centered AR is achieved via EEG signals acquired via the wireless and wearable \textit{Emotiv} headset, while facilitates naturalistic user behavior and can be employed for large-scale AR.

\subsection{Emotional impact of ads}
Ad-induced emotions have been shown to shape consumer behavior in a significant manner~\cite{Holbrook1984,Holbrook1987}. Although this key observation was made nearly three decades ago~\cite{Holbrook1987}, computational advertising methods till recently have matched low-level visual and semantic properties between video segments and candidate ads~\cite{videosense}. Recent work~\cite{Pham2013} indicates a shift form the traditional thinking by emphasizing that ad-evoked emotions can change brand perception among consumers. A very recent and closely related work to ours~\cite{Shukla2017acm} discusses how efficient affect recognition from ads via deep learning and multi-task learning can lead to improved online viewing experience. In this work, we show how effectively recognizing emotions from ads via content and user-based methods can achieve optimized insertion of ads onto streamed/broadcast videos via the CAVVA framework~\cite{cavva}. A user study shows that better ad AR translates to better ad memorability and enhanced user experience while watching an ad-embedded video stream.


\subsection{Computational advertising}
Exploiting affect recognition models for commercial applications has been a growing trend in recent years. The field of \textit{\textbf{computational advertising}} focuses on presenting contextually relevant ads to multimedia users for commercial benefits, social good or to induce behavioral change. Traditional computational advertising approaches hae worked by exclusively modeling low-level visual and semantic relevance between video scenes and ads~\cite{videosense}. A paradigm shift in this regard was introduced by the CAVVA framework, which proposed an optimization-based approach to insert ads onto a video stream based on the emotional relevance between the video scenes and candidate ads. CAVVA employed a \textit{content-centric} approach to match video scenes and ads in terms of emotional valence and arousal. However, this could be replaced by an interactive and \textbf{user-centric} framework as described in~ \cite{YadatiMMM2013}. We explore the use of both \textit{content-centric} (via CNN features) and \textbf{user-centric} (via EEG features) methods for formulating an ad-insertion strategy. A user study shows that CNN-based ad insertion results in better ad memorability, while an EEG-based strategy achieves the best user experience. The following section describes the compiled ad dataset, and the EEG acquisition protocol.

\section{Advertisement Dataset}\label{ad_set}
This section presents details regarding the ad dataset used in this study along with the protocol employed for collecting EEG responses for user-centric AR.

\subsection{Dataset Description}
\begin{figure}[t]
\includegraphics[width=0.32\linewidth]{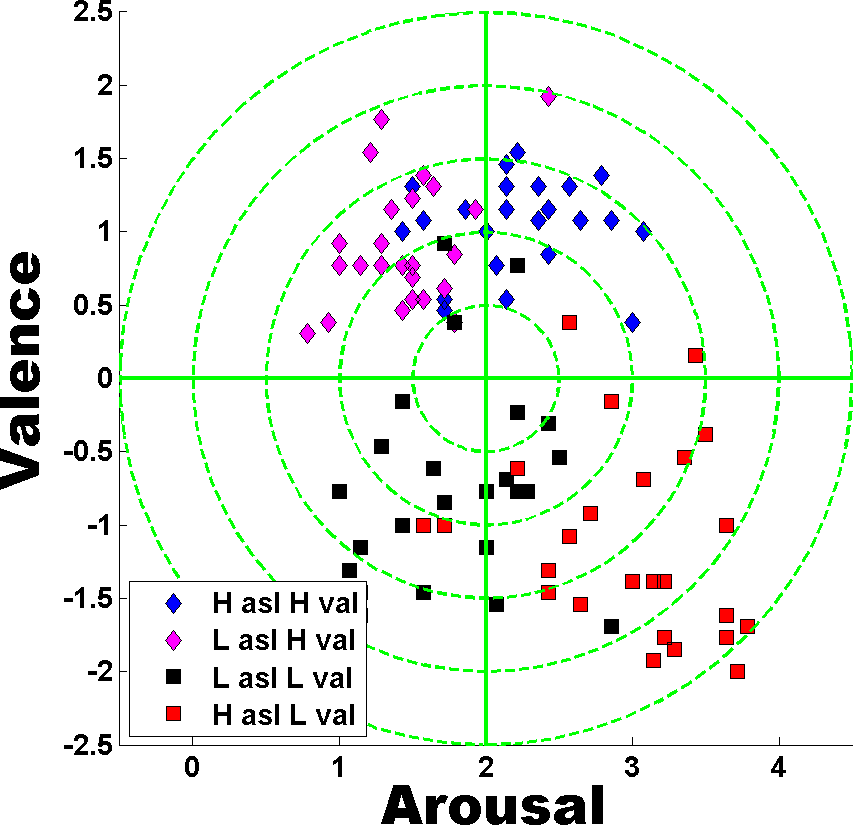}\hfill
\includegraphics[width=0.32\linewidth]{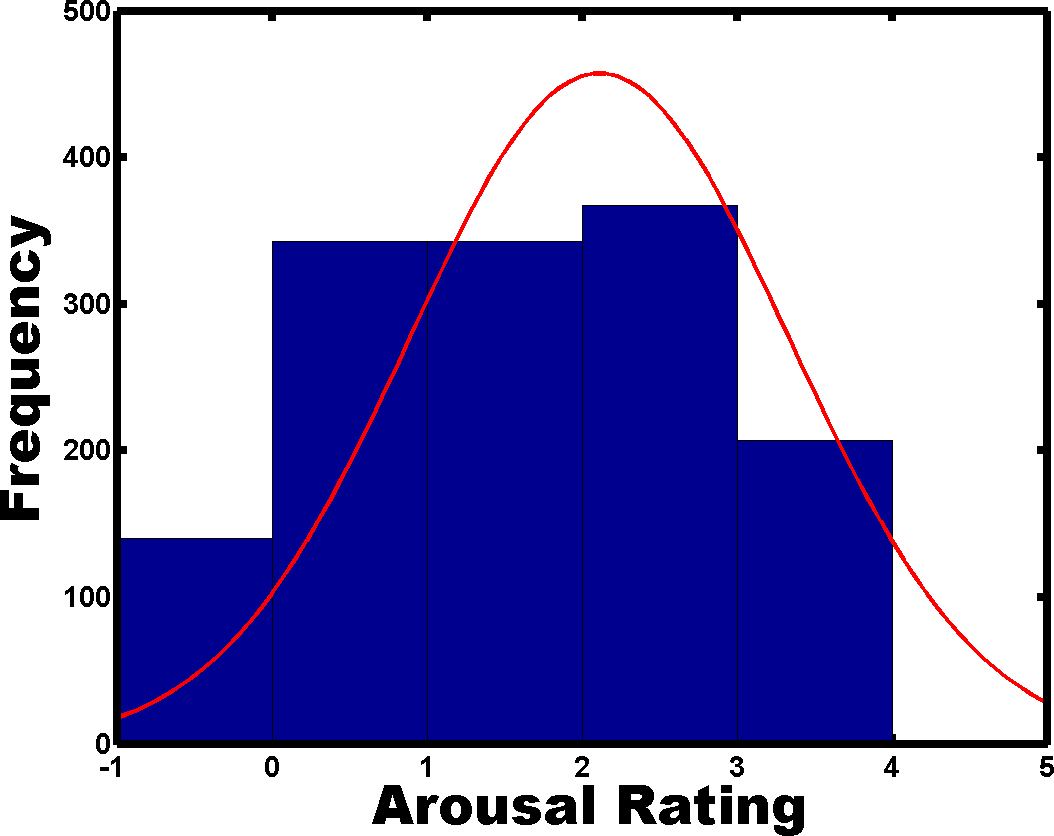}\hfill
\includegraphics[width=0.32\linewidth]{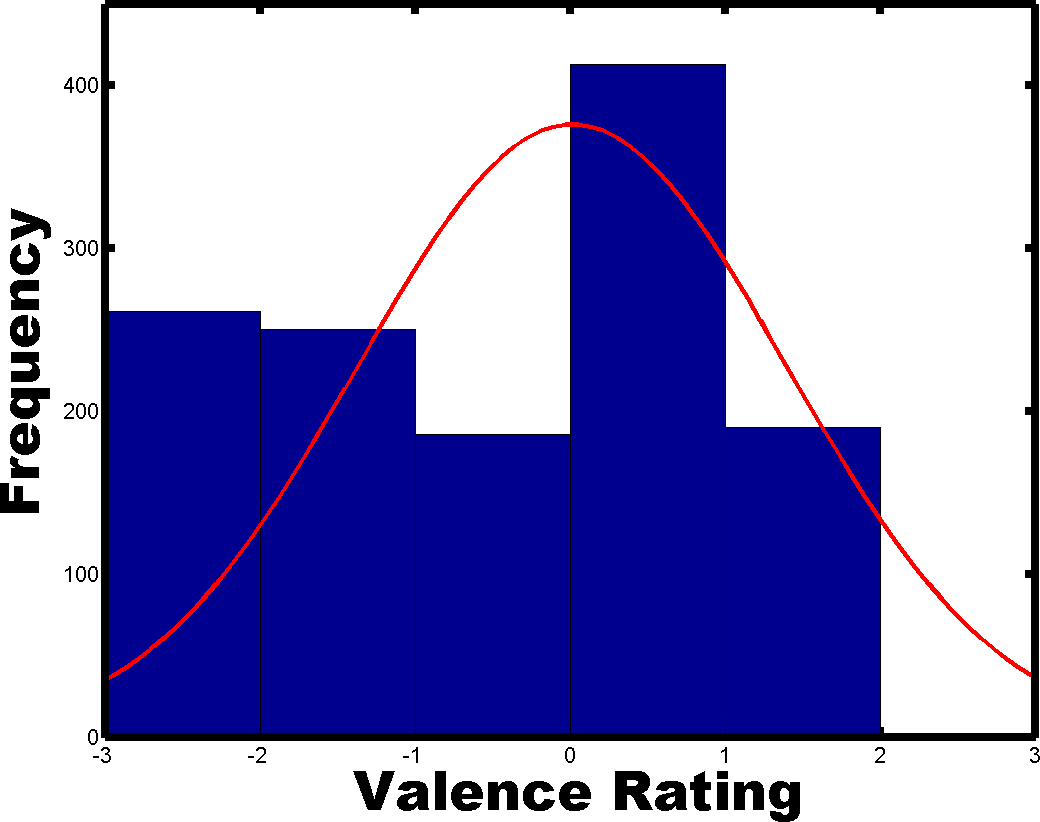}\vspace{-.1cm}
\caption{\label{Annot_dist} (left) Scatter plot of mean asl, val ratings color-coded with expert labels. (middle) Asl and (right) Val rating distribution with Gaussian pdf overlay (view under zoom).}\vspace{-.2cm}
\end{figure}

Defining \textbf{\textit{valence}} as the feeling of \textit{pleasantness}/\textit{unpleasantness} and \textbf{\textit{arousal}} as the \textit{intensity of emotional feeling} while viewing an audio-visual stimulus, five experts carefully compiled a dataset of 100, roughly 1-minute long commercial advertisements (ads) which are used in this work. These ads are publicly available\footnote{On video hosting websites such as YouTube.} and found to be uniformly distributed over the arousal--valence plane defined by Greenwald \textit{et al.}~\cite{greenwald1989} (Figure~\ref{Annot_dist}). An ad was chosen if there was consensus among all five experts on its valence and arousal labels (defined as either \textit{high} (H)/\textit{low} (L)). The high valence ads typically involved product promotions, while low valence ads were social messages depicting the ill effects of smoking, alcohol and drug abuse, \etc. Labels provided by experts were considered as \textbf{\textit{ground-truth}}, and used for all recognition experiments in this work. 
 
To evaluate the effectiveness of these ads as affective control stimuli, we examined how consistently they could evoke target emotions across viewers. To this end, the ads were independently rated by 14 annotators for valence (val) and arousal (asl)\footnote{Annotators were familiarized with emotional attributes prior to the rating task.}. All ads were rated on a 5-point scale, which ranged from -2 (\textit{very unpleasant}) to 2 (\textit{very pleasant}) for val and 0 (\textit{calm}) to 4 (\textit{highly aroused}) for asl. Table~\ref{tab:ads_des} presents summary statistics for ads over the four quadrants. Evidently, low val ads are longer and are perceived as more arousing than high val ads suggesting that they evoked stronger emotional feelings among viewers.

\begin{table}[t]
\vspace{-.1cm}
\fontsize{7}{7}\selectfont
\renewcommand{\arraystretch}{1.8}
\caption{\label{tab:ads_des} Summary statistics for quadrant-wise ads.}\vspace{-.2cm}
\centering
\begin{tabular}{|c|ccc|} \hline
\textbf{Quadrant} & \textbf{Mean length (s)} & \textbf{Mean asl} & \textbf{Mean val}  \\ \hline \hline
\textbf{H asl, H val} & 48.16 & 2.17 & \ 1.02 \\
\textbf{L asl, H val} & 44.18 & 1.37 & \ 0.91 \\
\textbf{L asl, L val} & 60.24 & 1.76 & -0.76 \\
\textbf{H asl, L val} & 64.16 & 3.01 & -1.16 \\ \hline
\end{tabular}
\vspace{-.3cm}
 \end{table}

Furthermore, we computed agreement among raters in terms of the (i) Krippendorff's $\alpha$ and (ii) Cohen's $\kappa$ scores. The $\alpha$ coefficient is applicable when multiple raters code data with ordinal scores-- we obtained $\alpha = 0.60$ and $0.37$ for val and asl implying valence impressions were most consistent across raters. We then computed the $\kappa$ agreement between annotator and ground-truth labels to determine concordance between the annotator and expert groups. To this end, we thresholded each rater's asl, val scores by their mean rating to assign H/L labels for each ad, and compared them against ground-truth labels. This procedure revealed a mean agreement of 0.84 for val and 0.67 for asl across raters. Computing $\kappa$ between the annotator and expert \textit{populations} by thresholding the mean asl, val score per ad across raters against the grand mean gave a $\kappa = 0.94$ for val and 0.67 for asl\footnote{Chance agreement corresponds to a $\kappa$ value of 0.}. Clearly, there is good-to-excellent agreement between annotators and experts on affective impressions with considerably higher concordance for val. The observed concordance between the independent expert and annotator groups affirms that the compiled 100 ads are effective control stimuli for affective studies.  

Another desirable property of an affective dataset is the independence of the asl and val dimensions. We (i) examined scatter plots of the annotator ratings, and (ii) computed correlations amongst those ratings. The scatter plot of the mean asl, val annotator ratings, and the distribution of asl and val ratings are presented in Fig.~\ref{Annot_dist}. The scatter plot is color-coded based on expert labels, and is interestingly different from the classical `C' shape observed with images~\cite{IAPS}, music videos~\cite{Koelstra} and movie clips~\cite{decaf} owing to the difficulty in evoking medium asl/val but strong val/asl responses. The distributions of asl and val ratings are also roughly uniform resulting in Gaussian fits with large variance, with modes observed at the median scale values of 2 and 0 respectively. A close examination of the scatter plot reveals that a number of ads are rated as moderate asl, but high/low val. This is owing to the fact that ads are designed to convey a strong positive or negative message to viewers, which is not typically true of images or movie scenes. Finally, Wilcoxon rank sum tests on annotator ratings revealed significantly different asl ratings for high and low asl ads ($p<0.00005$), and distinctive val scores for high and low valence ads ($p<0.000001$), consistent with expectation.    

Pearson correlation was computed between the asl and val dimensions with correction for multiple comparisons by limiting the false discovery rate to within 5\%~\cite{benjamini1995controlling}. This procedure revealed a weak and insignificant negative correlation of 0.19, implying that ad asl and val scores were largely uncorrelated. Overall, (i) Our ads constitute a control affective dataset as asl and val ratings are largely independent; (ii) Different from the `C'-shape characterizing the asl-val relationship for other stimulus types, asl and val ratings are uniformly distributed for the ad stimuli, and (iii) There is considerable concordance between the experts and annotators on affective labels, implying that the selected ads effectively evoke coherent emotions across viewers.

\subsection{EEG acquisition protocol} As 11 of the 14 annotators rated the ads for asl and val upon watching them, we acquired their Electroencephalogram (EEG) brain activations via the \textit{Emotiv} wireless headset. To maximize engagement and minimize fatigue during the rating task, these raters took a break after every 20 ads, and viewed the entire set of 100 ads over five sessions. Upon viewing each ad, the raters had a maximum of 10 seconds to input their asl and val scores via mouse clicks. The Emotiv device comprises 14 electrodes, and has a sampling rate of 128 Hz. Upon experiment completion, the EEG recordings were segmented into \textit{epochs}, with each epoch denoting the viewing of a particular ad. Upon removal of noisy epochs, we were left with a total of 804 \textit{clean} epochs. Each ad was preceded by a 1s fixation cross to orient user attention, and to measure resting state EEG power used for baseline power subtraction. The EEG signal was band-limited between 0.1--45 Hz, and independent component analysis (ICA) was performed to remove artifacts relating to eye movements, eye blinks and muscle movements. The following section describes the techniques employed for content and user-centered AR.

\section{Content \& User-centered Analysis}\label{Data_anal}

This section presents the modeling techniques employed for content-centered and user-centered ad affect recognition.

\subsection{Content-centered Analysis}
For content centered analysis, we employed a convolutional neural network (CNN)-based model, and the popular affective model of Hanjalic and Xu based on low-level audio visual descriptors~\cite{Hanjalic2005}. CNNs have recently become very popular for visual~\cite{alex12} and audio~\cite{Huang2014} recognition, but they require vast amounts of training data. As our ad dataset comprised only 100 ads, we fine-tuned the pre-trained \textit{places205}~\cite{alex12} model via the affective LIRIS-ACCEDE movie dataset~\cite{baveye2015liris}, and employed the fine-tuned model to extract emotional descriptors for our ads. This process is termed as \textit{\textbf{domain adaptation}} in machine learning literature.  

\begin{sloppypar}
In order to learn deep features for ad AR, we employed the \textit{Places205} CNN~\cite{Khosla2013} originally trained for image classification. \textit{Places205} is trained using the {Places-205} dataset comprising 2.5 million images involving 205 scene categories. The {Places-205} dataset contains a wide variety of scenes captured under varying illumination, viewpoint and field of view, and we hypothesized a strong relationship between scene perspective, lighting and the scene mood. The \textbf{LIRIS-ACCEDE} dataset contains asl, val ratings for $\approx$ 10 s long movie snippets, whereas our ads are about a minute-long with individual ads ranging from 30--120 s.
\end{sloppypar}

\subsubsection{FC7 Feature Extraction via CNNs}
For deep CNN-based ad AR, we represent the \textit{visual} modality using \textit{key-frame} images, and the \textit{audio} modality using \textit{spectrograms}. We fine-tune \textit{Places205} via the LIRIS-ACCEDE~\cite{baveye2015liris} dataset, and employ this model to compute the fully connected layer (fc7) visual and audio ad descriptors.

\paragraph*{Keyframes as Visual Descriptors}
From each video in the ad and LIRIS-ACCEDE datasets, we extract one \textit{key frame} every three seconds-- this enables extraction of a continuous video profile for affect prediction. This process generates a total of 1791 key-frames for our 100 ads.

\paragraph*{Spectrograms as Audio Descriptors}
Spectrograms (SGs) are visual representations of the audio frequency spectrum, and have been successfully employed for AR from speech and music~\cite{baveyethesis}. Specifically, transforming the audio content to a spectrogram image allows for audio classification to be treated as a visual recognition problem. We extract spectrograms over the 10s long LIRIS-ACCEDE clips, and consistently from 10s ad segments. This process generates 610 spectrograms for our ad dataset. Following~\cite{baveyethesis}, we combine multiple tracks to obtain a single spectrogram (as opposed to two for stereo). Each spectrogram is generated using a 40 ms window short time Fourier transform (STFT), with 20 ms overlap. Larger densities of high frequencies can be noted in the spectrograms for high asl ads, and these intense scenes are often characterized by sharp frequency changes. 

\paragraph*{CNN Training}
We use the Caffe~\cite{caffe} deep learning framework for fine-tuning \textit{places205}, with a momentum of 0.9, weight decay of 0.0005, and a base learning rate of 0.0001 reduced by $\frac{1}{10}^{th}$ every 20000 iterations. We totally train four binary classification networks to recognize high and low asl/val from audio/visual features. To fine-tune \textit{places205}, we use only the top and bottom 1/3rd LIRIS-ACCEDE videos in terms of asl and val rankings under the assumption that descriptors learned for the extreme-rated clips will effectively model affective concepts. 4096-dimensional \textit{fc7} layer outputs extracted from the four networks for our 100 ads are used in the experiments. 

\subsubsection{AR with audio-visual features}
We will mainly compare our CNN-based AR framework against the algorithm of Hanjalic and Xu~\cite{Hanjalic2005} in this work. Even after a decade, this algorithm remains one of the most popular AR baselines as noted from recent works such as~\cite{Koelstra,decaf}. In~\cite{Hanjalic2005}, asl and val are modeled via low-level descriptors describing motion activity, colorfulness, shot change frequency, voice pitch and sound energy in the scene. These hand-crafted features are intuitive and interpretable, and employed to estimate time-continuous asl and val levels conveyed by the scene. Table~\ref{tab:exp_det} summarizes the audio-visual features used for content-centric AR.

\subsection{User-centered analysis}
The 804 clean epochs obtained from the EEG acquisition process were used for user-centered analysis. However, these 804 epochs were of different lengths as the duration of each ad was variable. To maintain dimensional consistency, we performed user-centric AR experiments with (a) the \textit{first} 3667 samples ($\approx 30s$ of EEG data), (b) the \textit{last} 3667 samples and (c) the \textit{last} 1280 samples (10s of EEG data) from each epoch. Each epoch sample comprises data from 14 EEG channels, and the epoch samples were input to the classifier upon vectorization. 

\begin{table}[t]
\fontsize{6}{6}\selectfont
\renewcommand{\tabcolsep}{4.6pt}
\caption{Extracted features for content-centric AR.} \label{tab:exp_det} \vspace{-.2cm} 
\begin{center}                                                         
\begin{tabular}{@{}|c|ccc|@{}} 

\hline
\textbf{Attribute} & \multicolumn{3}{c|}{\textbf{Valence/Arousal}} \\ 

~ &{\textbf{Audio}}& {\textbf{Video}}&{\textbf{aud+vid (A+V)}}\\ \hline\hline
 
 {\textbf{CNN}} & 4096D Alexnet FC7   & 4096D Alexnet FC7 features by  & 8192D FC7 features \\

 \textbf{Features} & features obtained  & extracted from keyframes  &  with SGs + keyframes \\ 
        &  with 10s SGs.   & sampled every 3 seconds.            & over 10s intervals. \\ \hline

\textbf{Hanjalic~\cite{Hanjalic2005}} & Per-second sound  & Per-second shot change & Concatenation of \\ 
\textbf{Features} &  energy and pitch  & frequency and motion & audio-visual features. \\
 & statistics~\cite{Hanjalic2005}. & statistics~\cite{Hanjalic2005}. & ~ \\ \hline

\end{tabular}
\end{center}
\vspace{-.3cm}
\end{table}



\section{Experiments and Results}\label{ER}

\begin{table*}[htbp]
\fontsize{8}{8}\selectfont
\renewcommand{\arraystretch}{1.3}
\centering
\caption{Ad AR from content analysis. F1 scores are presented in the form $\mu \pm \sigma$. } \label{tab:cap} \vspace{-.2cm}
\begin{tabular}{|c|ccc|ccc|}
  \hline
 
	\multicolumn{1}{|c|}{\textbf{Method}} & \multicolumn{3}{c|}{\textbf{Valence}} & \multicolumn{3}{c|}{\textbf{Arousal}} \\ \hline 
	\multicolumn{1}{|c|}{~} & {\textbf{F1 (all)}} & {\textbf{F1 (L30)}} & {\textbf{F1 (L10)}}  & {\textbf{F1 (all)}} & {\textbf{F1 (L30)}} & {\textbf{F1 (L10)}}\\ \hline

	
	\textbf{Audio FC7 + LDA}  & 0.61$\pm$0.04 & 0.62$\pm$0.10 & 0.55$\pm$0.18 & 0.65$\pm$0.04 & 0.59$\pm$0.10 & {0.53$\pm$0.19}\\
	\textbf{Audio FC7 + LSVM} & 0.60$\pm$0.04 & 0.60$\pm$0.09 & 0.55$\pm$0.19 & 0.63$\pm$0.04 & 0.57$\pm$0.09 & 0.50$\pm$0.18\\
	\textbf{Audio FC7 + RSVM} & {0.64$\pm$0.04} & \textbf{0.66$\pm$0.08} & {0.62$\pm$0.17} & \textbf{0.68$\pm$0.04} & {0.60$\pm$0.10} & {0.53$\pm$0.19}\\ \hline

	\textbf{Video FC7 + LDA} & 0.69$\pm$0.02 & 0.79$\pm$0.08 & {0.77$\pm$0.13} & 0.63$\pm$0.03 & 0.58$\pm$0.10 & 0.57$\pm$0.18\\
	\textbf{Video FC7 + LSVM}& 0.69$\pm$0.02 & 0.74$\pm$0.08 & 0.70$\pm$0.15 & 0.62$\pm$0.02 & 0.57$\pm$0.09 & 0.52$\pm$0.17\\
	\textbf{Video FC7 + RSVM}& {0.72$\pm$0.02} & \textbf{0.79$\pm$0.07} & 0.74$\pm$0.15 & \textbf{0.67$\pm$0.02} & {0.62$\pm$0.10} & {0.58$\pm$0.19}\\ \hline
	
	\textbf{{A+V FC7 + LDA}} &  0.70$\pm$0.04 & 0.66$\pm$0.08 & 0.49$\pm$0.18 & 0.60$\pm$0.04 & 0.52$\pm$0.10 & {0.51$\pm$0.18}\\
	\textbf{A+V FC7 + LSVM}  &  0.71$\pm$0.04 & 0.66$\pm$0.07 & 0.49$\pm$0.19 & 0.56$\pm$0.04 & 0.49$\pm$0.10 & 0.47$\pm$0.19\\
	\textbf{A+V FC7 + RSVM}  &  \textbf{0.75$\pm$0.04} & {0.70$\pm$0.07} & {0.55$\pm$0.17} & \textbf{0.63$\pm$0.04} & {0.56$\pm$0.11} & 0.49$\pm$0.19\\ \hline
	
\textbf{{A+V Han + LDA}} & 0.59$\pm$0.09 &{0.63$\pm$0.08} & {0.64$\pm$0.12} & {0.54$\pm$0.09} & 0.50$\pm$0.10 & {0.58$\pm$0.08}\\
	\textbf{A+V Han + LSVM}  & {0.62$\pm$0.09} & {0.62$\pm$0.10} & {0.65$\pm$0.11} & 0.55$\pm$0.10 & {0.51$\pm$0.11} & 0.57$\pm$0.09\\
	\textbf{A+V Han + RSVM}  & \textbf{0.65$\pm$0.09} & {0.62$\pm$0.11} & 0.62$\pm$0.12 & \textbf{0.59$\pm$0.12} & {0.58$\pm$0.11} & 0.56$\pm$0.10\\ \hline
		
	\textbf{{A+V FC7 LDA DF}} & 0.60$\pm$0.04  &  {0.66$\pm$0.04}  &  {0.70$\pm$0.19} & 0.59$\pm$0.02 & 0.60$\pm$0.07 & {0.57$\pm$0.15} \\
	\textbf{A+V FC7 LSVM DF}  & {0.65$\pm$0.02}  &  {0.66$\pm$0.04}  &  0.65$\pm$0.08 & {0.60$\pm$0.04}  & {0.63$\pm$0.10} & {0.53$\pm$0.13} \\
	\textbf{A+V FC7 RSVM DF}  & \textbf{{0.72$\pm$0.04}}  &  {0.70$\pm$0.04}  &  {0.70$\pm$0.12} & {0.69$\pm$0.06} & \textbf{0.75$\pm$0.07} & {0.70$\pm$0.07}\\ \hline
	
\textbf{{A+V Han LDA DF}}   & 0.58$\pm$0.09 & 0.58$\pm$0.09 & \textbf{0.61$\pm$0.09} & {0.59$\pm$0.06} & {0.59$\pm$0.07} & {0.61$\pm$0.08}\\
	\textbf{A+V Han LSVM DF}  & 0.59$\pm$0.10 & 0.59$\pm$0.09 & 0.60$\pm$0.10 & {\textbf{0.61$\pm$0.05}} & {0.61$\pm$0.08} & {0.60$\pm$0.09}\\
	\textbf{A+V Han RSVM DF}  & 0.60$\pm$0.08 & 0.56$\pm$0.10 & {0.58$\pm$0.09} & {0.58$\pm$0.09} & {0.56$\pm$0.06} & {0.58$\pm$0.09}\\ \hline
	
  \hline
\end{tabular}
\vspace{0.1cm}
%
%
\fontsize{8}{8}\selectfont
\renewcommand{\arraystretch}{1.3}
\centering
\caption{Ad AR from EEG analysis. F1 scores are presented in the form $\mu \pm \sigma$. } \label{tab:uscap} \vspace{-.2cm}
\begin{tabular}{|c|ccc|ccc|}
  \hline
	\multicolumn{1}{|c|}{\textbf{Method}} & \multicolumn{3}{c|}{\textbf{Valence}} & \multicolumn{3}{c|}{\textbf{Arousal}} \\ \hline 
	\multicolumn{1}{|c|}{~} & {\textbf{F1 (F30)}} & {\textbf{F1 (L30)}} & {\textbf{F1 (L10)}}  & {\textbf{F1 (F30)}} & {\textbf{F1 (L30)}} & {\textbf{F1 (L10)}}\\ \hline

	
	\textbf{LDA} & 0.79 $\pm$ 0.03 & 0.79 $\pm$ 0.03 & 0.75 $\pm$ 0.03 & 0.75 $\pm$ 0.03 & 0.74 $\pm$ 0.03 & 0.71 $\pm$ 0.04\\
	\textbf{LSVM} & 0.77 $\pm$ 0.03 & 0.76 $\pm$ 0.04 & 0.77 $\pm$ 0.05 & 0.74 $\pm$ 0.03 & 0.73 $\pm$ 0.02 & 0.69 $\pm$ 0.04\\
	\textbf{RSVM} & \textbf{{0.83 $\pm$ 0.03}} & \textbf{0.83 $\pm$ 0.03} & \textbf{{0.81 $\pm$ 0.03}} & \textbf{0.80 $\pm$ 0.02} & \textbf{{0.80 $\pm$ 0.03}} & \textbf{{0.76 $\pm$ 0.04}}\\ \hline
\end{tabular}
\vspace{-.2cm}
\end{table*}

We first provide a brief description of the classifiers used and settings employed for  binary content-centric and user-centric AR, where the objective is to assign  a binary (H/L) label for asl and val evoked by each ad, using the extracted fc7/low-level audio visual/EEG  features. The ground truth here is provided by the experts, and has a substantial agreement with the user ratings in Sec. 3.1. Experimental results will be discussed thereafter.

\paragraph*{Classifiers:} We employed the Linear Discriminant Analysis (LDA), linear SVM (LSVM) and Radial Basis SVM (RSVM) classifiers in our AR experiments. LDA and LSVM separate H/L labeled training data with a hyperplane, while RSVM is a non-linear classifier which separates H and L classes, linearly inseparable in the input space, via transformation onto a high-dimensional feature space. 

\paragraph*{Metrics and Experimental Settings:} We used the F1-score (F1), defined as the harmonic mean of precision and recall as our performance metric, due to the unbalanced distribution of positive and negative samples. For content-centric AR, apart from unimodal (audio (A) or visual (V)) fc7 features, we also employed feature fusion and probabilistic decision fusion of the unimodal outputs. Feature fusion (A+V) involved concatenation of fc7 A and V features over 10 s windows (see Table~\ref{tab:exp_det}), while the $W_{est}$ technique~\cite{koelstra2012fusion} was employed for decision fusion (DF). In DF, the test label is assigned the index $i$ corresponding to maximum $P_i = \sum_{i=1}^2 \alpha_i^*t_ip_i$, where $i$ denotes the A,V modalities, $p_i$'s denote posterior A,V classifier probabilities and $\{\alpha_i^*\}$ are the optimal weights maximizing test F1-score, and determined via a 2D grid search. If $F_i$ denotes the training F1-score for the $i^{th}$ modality, then $t_i = \alpha_i F_i/\sum_{i=1}^2 \alpha_i F_i$ for given $\alpha_i$. Note that the use of a validation set for parameter tuning is precluded by the small dataset size as with [1,18] and that the DF results denote 'maximum possible' performance.

As the Hanjalic (Han) algorithm~\cite{Hanjalic2005} uses audio plus visual features to model asl and val, we only consider (feature and decision) fusion performance in this case. User-centered AR uses only EEG information. As we evaluate AR performance on a small dataset, AR results obtained over 10 repetitions of 5-fold cross validation (CV) (total of 50 runs) are presented. CV is typically used to overcome the \textit{overfitting} problem on small datasets, and the optimal SVM parameters are determined from the range $[10^{-3},10^{3}]$ via an inner five-fold CV on the training set. Finally, in order to examine the temporal variance in AR performance, we present F1-scores obtained over (a) all ad frames (`All'), (b) last 30s (L30) and (c) last 10s (L10) for \textit{content-centered} AR, and (a) first 30s (F30), (b) last 30s (L30) and (c) last 10s (L10) for \textit{user-centered} AR. These settings were chosen bearing in mind that EEG sampling rate is much higher than the audio or video sampling rate.      

\subsection{Results Overview}
Tables~\ref{tab:cap} and \ref{tab:uscap} respectively present content-centric and user-centric AR results for the various settings described above. The highest F1 score achieved for a given temporal setting across all classifiers and either unimodal or multimodal features is denoted in bold. Based on the observed results, we make the following claims. 

Superior val recognition is achieved with both \textit{content-centric} and \textit{user-centric} methods. Focusing on \textit{content-centric} results, \textbf{\textit{unimodal fc7 features}}, val (peak F1 = 0.79) is generally recognized better than asl (peak F1 = 0.68) and especially with video features. A and V fc7 features perform comparably for asl. Concerning recognition with \textbf{\textit{fused fc7 features}}, comparable or better F1 scores are achieved with multimodal approaches. In general, better recognition is achieved via decision fusion as compared to feature fusion\footnote{To our knowledge, either of feature or decision fusion may work better depending on the specific problem and available features.}. For val, the best fusion performance (0.75 with feature fusion and RSVM classifier) is superior compared to A-based (F1 = 0.66), but inferior compared to V-based (F1 = 0.79) recognition. Contrastingly for asl, fusion F1-score (0.75 with DF) considerably outperforms unimodal methods (0.68 with A, and 0.67 with V). 
Comparing \textit{\textbf{A$+$V fc7 vs Han}} features, fc7 descriptors clearly outperform Han features and the difference in performance is prominent for val, while comparable recognition is achieved with both features for asl. The RSVM classifier produces the best F1-scores for both asl and val with unimodal and multimodal approaches. 

\textit{User-centric} or \textbf{\textit{EEG}}-based AR results are generally better than \textit{content-centric} results achieved under similar conditions. The best \textit{user-centric} val and asl F1-scores are considerably higher than the best \textit{content-centric} results. Again, val is recognized better than asl with EEG data (as with the \textit{content-centric} case), which is interesting as EEG is known to correlate better with asl rather than val. Nevertheless, positive val is found to correlate with higher activity in the frontal lobes as compared to negative val as noted in~\cite{oude2006eeg}, and the Emotiv device is known to efficiently capture frontal lobe activity despite its limited spatial resolution. Among the three classifiers considered with EEG data, RSVM again performs best while LSVM performs worst. 

Focusing on the different temporal conditions considered in our experiments, relatively small $\sigma$ values are observed for the `All' \textit{content-centric} condition with the five-fold CV procedure (Table~\ref{tab:cap}), especially with fc7 features. Still lower $\sigma$'s can be noted with EEG-based classification results, suggesting that our overall AR results are \textit{\textbf{minimally impacted}} by \textit{\textbf{overfitting}}. Examining \textit{\textbf{temporal windows}} considered for \textit{content-centered} AR, higher $\sigma$'s are observed for the L30 and L10 cases, which denote model performance on the terminal ad frames. Surprisingly, one can note a general degradation in asl recognition for the L30 and L10 conditions with A/V features, while val F1-scores are more consistent. 

Three inferences can be made from the above observations, namely, (1) Greater heterogeneity in the ad content towards endings is highlighted by the large variance with fusion approaches; (2) Fusion models synthesized with Han features appear to be more prone to overfitting, given the generally larger $\sigma$ values seen with the models; (3) That asl recognition is lower in the L30 and L10 conditions highlights the limitation of using a \textit{single} asl/val label (as opposed to dynamic labeling) over time. Generally lower F1-scores achieved for asl with all methods suggests that asl is a more transient phenomenon as compared to val, and that coherency between \textit{content-based} val features and labels is sustainable over time.           

\textit{User-centered} AR results obtained over the first 30, last 30 and final 10 s for the ads are relatively more stable than \textit{content-centered} results, especially for val. However, there is a slight dip in AR performance for asl over the final 10s. As the ads were roughly one minute long, we can infer that (a) the consistent F1 scores achieved for the firs and last 30s suggests that humans tend to perceive the ad mood rather quickly. This is in line with the objective of ad makers, who endeavor to convey an effective message within a short time duration. However, the dip in asl performance over the final 10s as with \textit{content centered} methods again highlights the limitation of using a single affective label over the entire ad duration.

\subsection{Discussion} We now summarize and compare the \textit{content-centric} and \textit{user-centric} AR results. Between the \textit{content-centric} features, the deep CNN-based \textit{\textbf{fc7}} descriptors considerably outperform the audio-visual \textit{\textbf{Han}} features. Also, the classifiers trained with \textit{\textbf{Han}} features are more prone to over-fitting than \textit{\textbf{fc7}}-based classifiers, suggesting that the CNN descriptors are more robust as compared to low-level \textit{\textbf{Han}} descriptors. Fusion-based approaches do not perform much better than unimodal methods. However, \textbf{\textit{EEG}}-based AR achieves the best performance, considerably outperforming content-based features and thereby endorsing the view that emotions are best characterized by human behavioral cues.

Superior val recognition is achieved with both \textit{content-centric} and \textit{user-centric} AR methods. Also, temporal analysis of classification results reveals that content-based val features as well as user-based val impressions are more stable over time, but asl impressions are transient. Cumulatively, the obtained results highlight the need for fine-grained and dynamic AR methods as against most contemporary studies which assume a single, \textit{static} affective label per stimulus.

\section{Computational Advertising- User Study}\label{US}
\input{user_study}

\section{Discussion and Conclusion}\label{CFW}
This work evaluates the efficacy of content-centric and user-centric techniques for ad affect recognition. At the outset, it needs to be stressed that content and user-centered AR methods encode complementary emotional information. Content-centric approaches typically look for emotional cues from low-level audio-visual (or textual) features, and do not include the human user as part of the computational loop; recent developments in the field of CNNs~\cite{alex12} have now made it possible to extract high-level emotion descriptors. Nevertheless, emotion is essentially a human feeling, and best manifests via user behavioral cues (\eg, facial emotions, speech and physiological signals), which explains why a majority of contemporary AR methods are user-centered~\cite{Koelstra,Zheng2014,decaf}. With the development of affordable, wireless and wearable sensing technologies such as \textit{Emotiv}, AR from large scale user data (termed \textit{crowd modeling}) is increasingly becoming a reality. 

We specifically evaluate the performance of two content-centered methods, the popular \textbf{Han} baseline for affect prediction from low-level audio-visual descriptors, and a \textbf{Deep} CNN-based framework which learns high-dimensional emotion descriptors from video frames or audio spectrograms, against the user-centered approach which employs \textbf{EEG} brain responses acquired from eleven users for AR. Experimental results show that while the deep CNN framework outperforms the Han method, it nevertheless performs inferior to an SVM-based classifier trained on EEG epochs for asl and val recognition. A study involving 12 users to examine if improved AR facilitates computational advertising reveals that (1) Ad memorability is maximized with better modeling of the ad affect via the Deep and EEG methods, and (2) Viewing experience is also enhanced by better matching of affective scores among the ads and video scenes. To our knowledge, this paper represents the first affective computing work to establish a direct relationship between objective AR performance and subjective viewer opinion.

Future work will focus on the development on effective alternative strategies to CAVVA for video-in-video advertising, as CAVVA is modeled on \textit{ad-hoc} rules derived from consumer psychology literature. Also, we observe that EEG-encoded affective information is complementary to representations learned by the Han and Deep CNN approaches, as EEG signals are derived from human users and there is little correlation between the val scores computed via the content and user-centered methods (Sec.~\ref{US-AdIns}). This reveals the potential for fusion strategies where \textit{content-centric} and \textit{user-centric} cues can be fused in a cross-modal decision making framework, as successfully attempted in prior~\cite{nusef2010,cntxtarxiv,retar2013} problems.

\begin{acks}
This research is supported by the National Research
Foundation, Prime Ministers Office, Singapore under its
International Research Centre in Singapore Funding Initiative.
\end{acks}

\bibliographystyle{ACM-Reference-Format}
\balance
\bibliography{affect_ads_icmi} 

\end{document}

%% file: user_study.tex
Given that superior ad AR is achieved with user EEG responses (see Table \ref{tab:uscap}), we examined if enhanced AR resulted in the insertion of appropriate ads at vantage temporal positions within a streamed video, as discussed in the CAVVA video-in-video ad insertion framework~\cite{cavva}. CAVVA is an optimization-based framework for ad insertion onto streamed videos (as with \textit{YouTube}). It formulates an advertising schedule by modeling the emotional relevance between video scenes and candidate ads to determine (a) the subset of ads for insertion, and (b) the temporal positions (typically after a scene ending) at which the chosen ads are to be inserted. In effect, CAVVA aims to strike a balance between (a) maximizing ad impact in terms of brand memorability, and (b) minimally disrupting (or enhancing) viewer experience while watching the program video onto which ads are inserted. We hypothesized that better ad affect recognition should lead to optimal ad insertions, and consequently better viewing experience. To this end, we performed a user study to compare the subjective quality of advertising schedules generated via ad asl and val scores generated with the content-centric \textbf{Han}~\cite{Hanjalic2005} and \textbf{Deep} CNN models, and the user-centric \textbf{EEG} model. 

\subsection{Dataset}\label{US-DS} 

For performing the user study, we used 28 ads (out of the 100 in the original dataset), and three program videos. The ads were equally divided into four quadrants of the valence-arousal plane based on asl and val labels provided by experts. The program videos were scenes from a television sitcom (\textit{friends}) and two movies (\textit{ipoh} and \textit{coh}), which predominantly comprised social themes and situations capable of invoking high-to-low valence and moderate arousal (see Table~\ref{tab:progdetails} for summary statistics). Each of the program videos comprised eight scenes implying that there were seven candidate ad-insertion points in the middle of each sequence. The average scene length was found to be 118 seconds.

\begin{table}[t]
\fontsize{8}{8}\selectfont
\renewcommand{\arraystretch}{1.3}
\centering
\caption{Summary of program video statistics.}\vspace{-.2cm}
\begin{tabular}{|c|c|cc|}
  \hline
	\multicolumn{1}{|c|}{\textbf{Name}} & \multicolumn{1}{c|}{\textbf{Scene length (s)}} & \multicolumn{2}{c|}{\textbf{Manual Rating}} \\
	 \hline 
	\multicolumn{1}{|c|}{~} & {~} & {\textbf{Valence}} & {\textbf{Arousal}}\\ \hline
	\textbf{coh}  & 127$\pm$46 & {0.08$\pm$1.18} & {1.53$\pm$0.58}\\
	\textbf{ipoh}  & 110$\pm$44 & {0.03$\pm$1.04} & {1.97$\pm$0.49}\\
	\textbf{friends}  & 119$\pm$69 & {1.08$\pm$0.37} & {2.15$\pm$0.65}\\
\hline
\end{tabular}
\vspace{-.2cm}
\label{tab:progdetails}
\end{table}

\subsection{Advertisement insertion strategy}~\label{US-AdIns}
We used the three aforementioned models to perform ad affect estimation. For the 24 program video scenes (3 videos $\times$ 8 scenes), the average of asl and val ratings acquired from three experts was used to denote affective scores. For the ads, affective scores were computed as follows. For the \textbf{Deep} method, we used normalized softmax class probabilities~\cite{bishop:2013} output by the video-based CNN model for val estimation, and probabilities from the audio CNN for asl estimation. The mean score over all video/audio ad frames was used to denote the affective score in this method. The average of the per-second asl and val level estimates over the ad duration was used to denote affective scores for the \textbf{Han} approach. Mean of SVM class posteriors over all EEG epochs was used for the \textbf{EEG} method. We then adopted the CAVVA optimization framework~\cite{cavva} to obtain nine unique \textbf{video program sequences} (with average length of 19.6 minutes) comprising the inserted ads. These video program sequences comprised ads inserted via the three affect estimation approaches onto each of the three program videos. Exactly 5 ads were inserted (out of 7 possible) onto each program video. 21 of the 28 chosen ads were inserted at least once into the nine video programs, with maximum and mean insertion frequencies of 5 and 2.14 respectively.

\subsection{Experiment and Questionnaire Design}~\label{US-Ques}

To evaluate the subjective quality of the generated video program sequences and thereby the utility of the three affect estimation techniques for computational advertising, we recruited 12 users (5 female, mean age 19.3 years) who were university undergraduates/graduates. Each of these users viewed a total of three {video program sequences}, corresponding to the three program videos with ad insertions performed using \textit{one} of the three affect estimation approaches.  We used a randomized 3$\times$3 Latin square design in order to cover all the nine generated sequences with every set of three users. Thus, each {video program sequence} was seen by four of the 12 viewers, and we have a total of 36 unique responses.

We designed a questionnaire for the user evaluation so as to reveal whether the generated video program sequences (a) included seamless ad insertions, (b) facilitated user engagement (or alternatively, resulted in minimum disruption) towards the streamed video and inserted ads and (c) ensured good overall viewer experience. To this end, we evaluated whether a particular ad insertion strategy resulted in (i) increased brand recall (both \textit{immediate} and \textit{day-after} recall) and (ii) minimal viewer disturbance or enhanced user viewing experience. 

Recall evaluation to intended to verify if the inserted ads were attended to by viewers, and the immediate and day-after recall were \textit{\textbf{objective}} measures that quantified the impact of ad insertion on the short-term (immediate) and long-term (day-after) memorability of advertised content, upon viewing the program sequences. Specifically, we measured the proportion of (i) inserted ads that were recalled correctly (\textit{Correct} recall), (ii) inserted ads that were not recalled (\textit{Forgotten}) and (iii) non-inserted ads incorrectly recalled as viewed (\textit{Incorrect} recall). For those inserted ads which were correctly recalled, we also assessed whether viewers perceived them to be contextually (emotionally) relevant to the program content.

Upon viewing a video program sequence, the viewer was provided with a representative visual frame from each of the 28 ads  to test ad recall along with a sequence-specific response sheet. In addition to the recall related questions, we asked viewers to indicate if they felt that the recalled ads were inserted at an appropriate position in the video (\textit{Good insertion}) to verify if ad positioning positively influenced recall. All recall and insertion quality-related responses were acquired from viewers as binary values. In addition to these objective measures, we defined a second set of \textit{\textbf{subjective}} user experience measures, and asked users to provide ratings on a Likert scale of 0--4 for the following questions with 4 implying \textit{best} and 0 denoting \textit{worst}:

\begin{itemize}[noitemsep,nolistsep]
\item[1.]Were the advertisements uniformly distributed across the video program?
\item[2.]	Did the inserted advertisements blend with the program flow?
\item[3.]	Whether the inserted ads were relevant to the surrounding scenes with respect to their content and mood?
\item[4.]	What was the overall viewer experience while watching each video program?
\end{itemize}

Each participant filled the recall and experience-related questionnaires immediately after watching each video program. Viewers also filled in the day-after recall questionnaire, a day after completing the experiment.

\begin{figure*}[t]
\centerline{\includegraphics[width=0.33\linewidth,height=3cm]{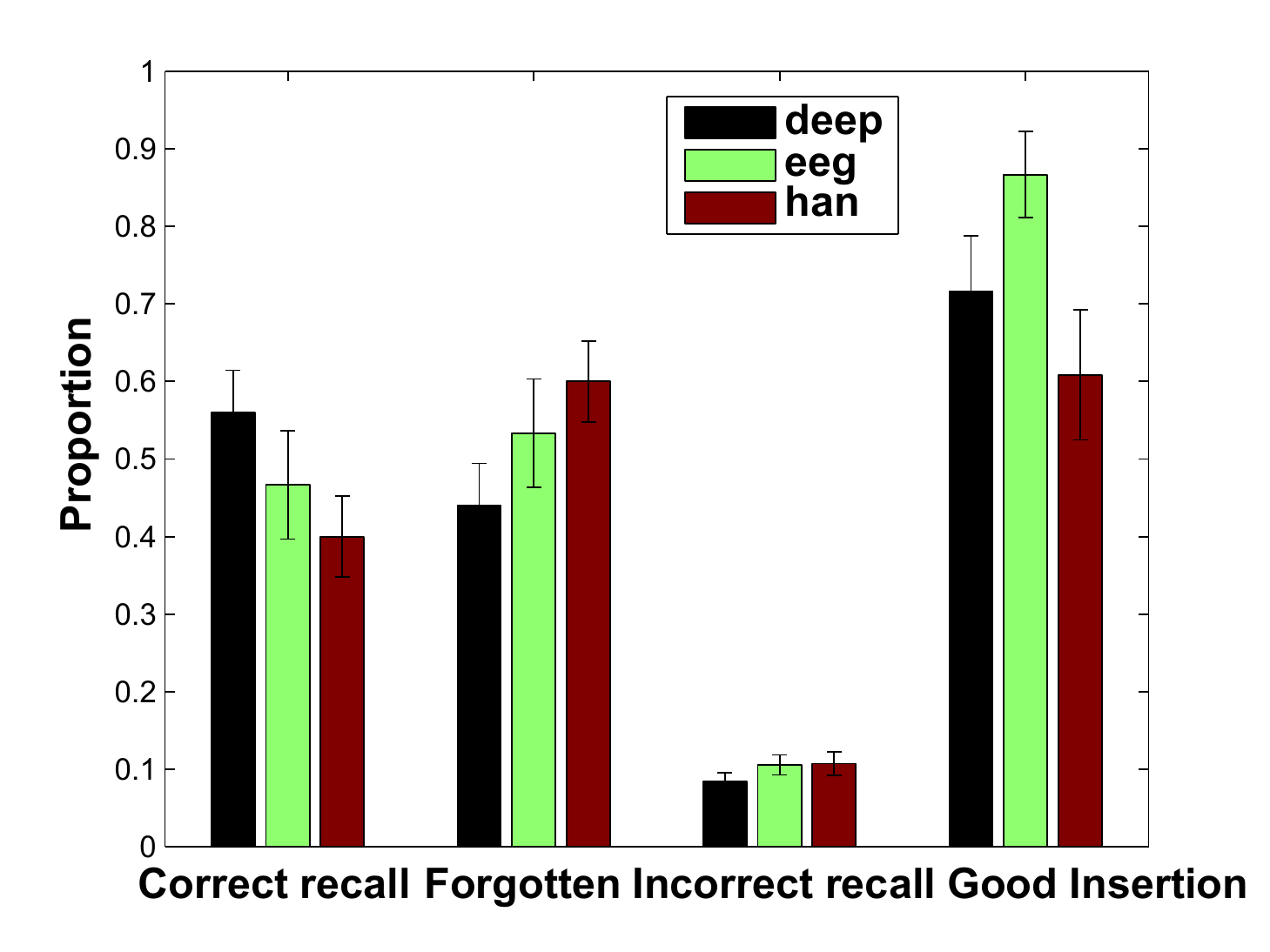}\hspace{0.05cm}
\includegraphics[width=0.33\linewidth,height=3cm]{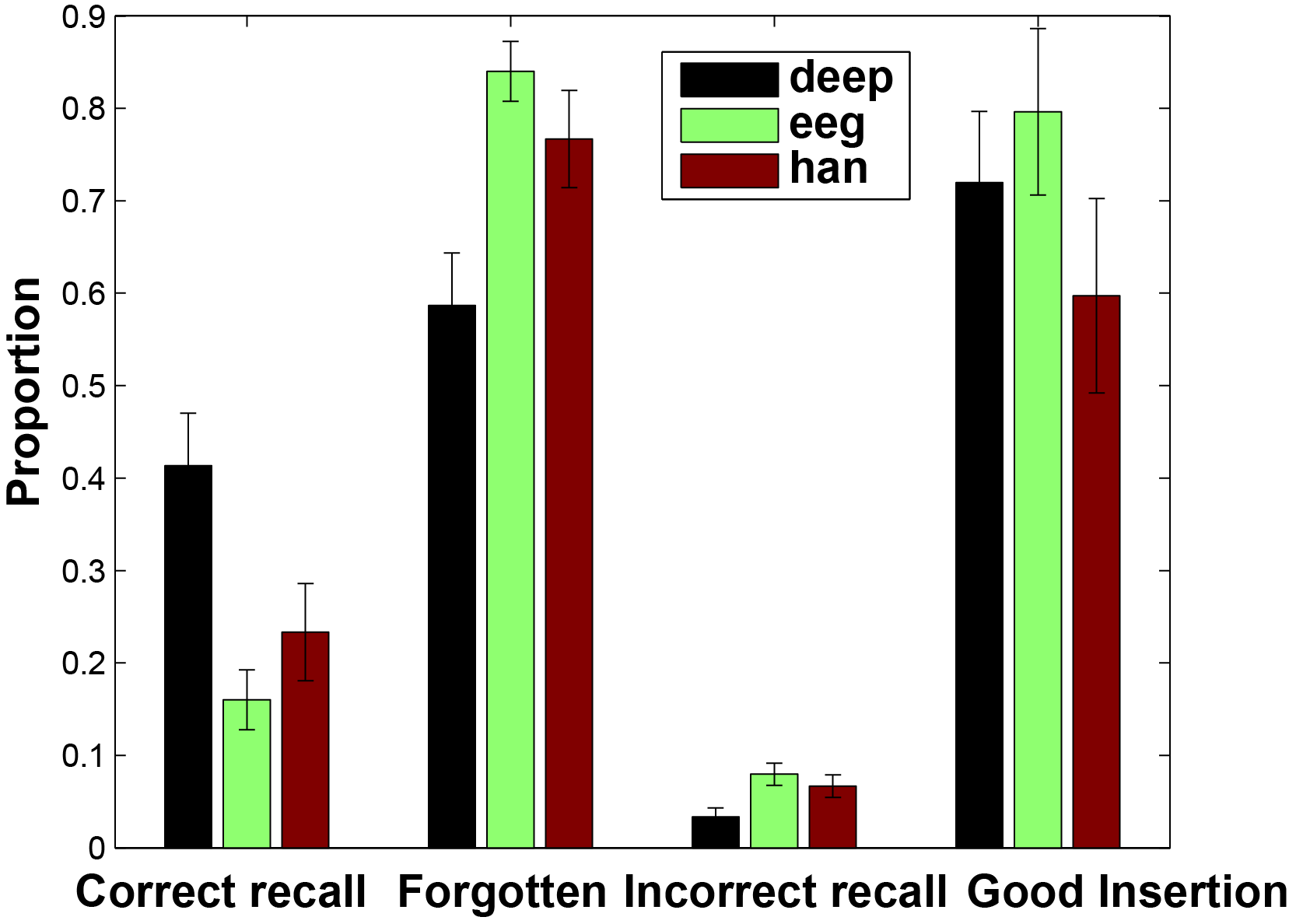}\hspace{0.05cm}
\includegraphics[width=0.33\linewidth,height=3cm]{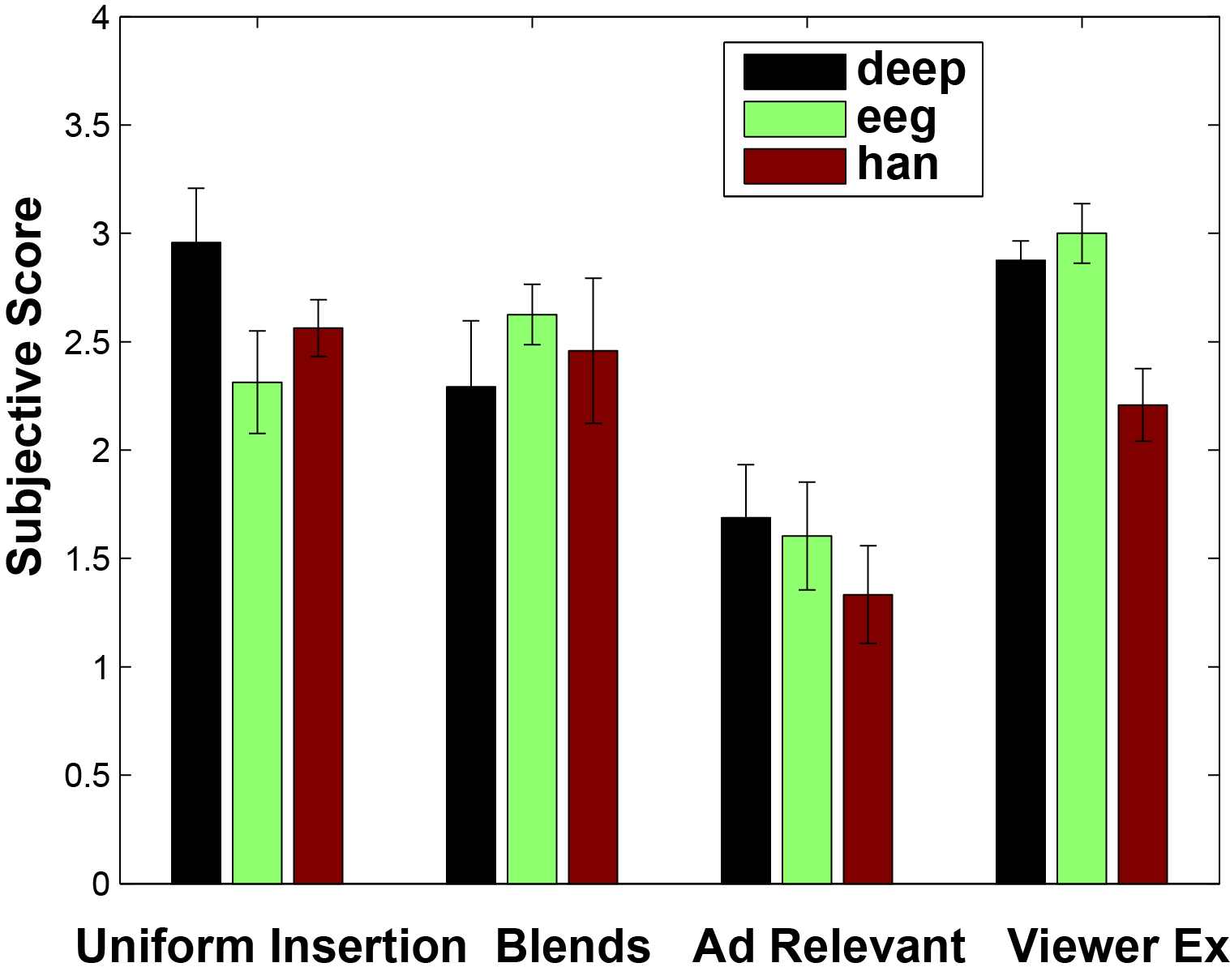}}
\centerline{\textbf{Immediate Recall}\hspace{0.2\linewidth}\textbf{Day-after recall}\hspace{0.23\linewidth}\textbf{User Experience}} \vspace{-.2cm}
\caption{\label{US_results} Summary of user study results in terms of recall and user experience-related measures. Error bars denote unit standard deviation.}
\vspace{-.2cm}
\end{figure*}

\subsection{Results and Discussion}


As mentioned previously, scenes from the program videos were assigned asl, val scores based on manual ratings from three experts, while the Deep, Han and EEG-based methods were employed to compute affective scores for ads. The overall quality of the CAVVA-generated video program sequence hinges on the quality of affective ratings assigned to both the video scenes and ads. In this regard, we hypothesized that better ad affect estimation would result in optimized ad insertions. 

Firstly, we computed the similarity in terms of the ad asl and val scores generated by the three approaches in terms of Pearson correlations, and found that (1) there was significant and positive correlation between asl scores generated by the Han--EEG ($\rho = 0.5, p<0.01$) as well as the Han--Deep methods ($\rho = 0.42, p<0.05$). However, the Deep and EEG-based asl scores did not agree significantly ($\rho = 0.22, \text{n.s.}$). For val, the only significant correlation was noted between the Han and Deep approaches ($\rho = 0.41, p<0.05$), while the Han and EEG ($\rho = 0.07, \text{n.s.}$) as well as the Deep and EEG val scores ($\rho = 0.20, \text{n.s.}$) were largely uncorrelated. This implies that while methods content-centric and user-centric methods agree well on asl scores, there is significant divergence between the val scores generated by the two approaches.

Based on the questionnaire responses received from viewers, we computed the mean proportions for correct recall, ad forgottenness, incorrect recall and good insertions immediately and a day after the experiment. Figure~\ref{US_results} presents the results of our user study and there are several interesting observations. A key measure indicative of a successful advertising strategy is \textit{high brand recall}~\cite{Holbrook1984, cavva, YadatiMMM2013}, and the immediate and day-after recall rates observed for three considered approaches are presented in Fig.~\ref{US_results} (left),(middle). Video program sequences obtained with {Deep} affective scores result in high immediate and day-after recall, least ad forgottenness and least incorrect recall. Ads inserted via the EEG method are found to be the best inserted, even if they have relatively lower recall rates as compared to the Deep approach ($p < 0.05$ for independent $t$-test). Ads inserted via Han-generated affective scores have the least immediate recall and are also forgotten the most, and are also perceived as the the worst inserted. The trends observed for immediate and day-after recall are slightly different, but the various recall measures are clearly worse for the day-after condition with a very high proportion of ads being forgotten. Nevertheless, the observed results clearly suggest that the Deep and EEG approaches which achieve superior AR compared to the Han method also lead to better ad memorability. 

However, it needs to be noted that higher ad recall does not directly translate to a better viewing experience. On the contrary, some ads may well be remembered because they disrupted the program flow and distracted viewers. In order to examine the impact of the affect-based ad insertion strategy on viewing experience, we computed the mean subjective scores acquired from users (Fig.~\ref{US_results}(right)). Here again, the Deep method scores best in terms of uniform insertion and ad relevance, while the EEG method performs best with respect to blending and viewer experience ($p < 0.05$ with two-sample $t$-tests in all cases). Interestingly, the Han method again performs worst in terms of ad relevance and viewer experience. The CAVVA optimization framework~\cite{cavva} has two components-- one for selection of ad insertion points into the program video, and another for selecting the set of ads to be inserted. Asl scores only play a role in the choice of insertion points, whereas val scores influence both components. In this context, the two best methods for val recognition, which also outperform the Han approach for asl recognition,  maximize both ad recall and viewing experience.